\documentclass[authoryear,12pt]{elsarticle}

\usepackage{enumitem}
\usepackage{graphicx}
\usepackage{amssymb}
\usepackage{amsthm}
\usepackage{amsfonts}
\usepackage{amsmath}
\usepackage[figuresright]{rotating}
\usepackage{enumerate}
\usepackage[latin1]{inputenc}
\usepackage{tabularx}
\usepackage{natbib}
\usepackage{breakcites}
\usepackage{pifont}
\usepackage{geometry}
\usepackage{txfonts}
\usepackage{color}
\definecolor{slateblue}{RGB}{106,90,205}
\usepackage[bookmarksnumbered=true,bookmarksopen=false, hyperfootnotes=true,  colorlinks=true, linkcolor=slateblue,citecolor=slateblue,urlcolor=slateblue, plainpages=false]{hyperref}
\usepackage{threeparttable}
\usepackage{array}
\usepackage{multirow}
\usepackage{dcolumn}
\usepackage[table]{xcolor}
\usepackage{framed}
\usepackage{xcolor}
\usepackage{natbib}
\usepackage{epstopdf}
\usepackage[capposition=top]{floatrow}
\usepackage{mathrsfs}
\usepackage{lscape}
\usepackage{setspace}
\usepackage{subcaption}
\usepackage{booktabs}
\usepackage{tikz} 
\usetikzlibrary{shapes,arrows} 
\usepackage{accents}
\usepackage{pdflscape}
\usepackage{todonotes}
\usepackage{soul}
\usepackage{makecell}

\soulregister\citet7
\soulregister\citep7
\soulregister\ref7
\soulregister\pageref7

\newcommand{\expectation}[3][0]{%
  \ifcase#1
     E( #2 \mid #3 )
     \or E \bigl( #2 \bigm\vert #3 \bigr)
     \or E \Bigl( #2 \Bigm\vert #3 \Bigr)
     \or E \biggl( #2 \biggm\vert #3 \biggr)
     \or E \Biggl( #2 \Biggm\vert #3 \Biggr)
  \else
     E \left( #2  \;\middle\vert\; #3 \right)
  \fi}

\usepackage{lscape} 

\makeatletter
\def\ps@pprintTitle{%
 \let\@oddhead\@empty
 \let\@evenhead\@empty
 \def\@oddfoot{}%
 \let\@evenfoot\@oddfoot}
\makeatother

\begin{document}

\begin{frontmatter}

\title{An ontological investigation of unimaginable events}

\author{Thomas Santoli\fnref{fn1}}

\author{Christoph Siebenbrunner\fnref{fn1}}

\ead{christoph.siebenbrunner@maths.ox.ac.uk}

\fntext[fn1]{University of Oxford, Mathematical Institute. All views expressed herein are those of the authors and do not necessarily reflect the views of any affiliating organization.}

\date{\today}

\begin{abstract}
We show that, under mild assumptions, some unimaginable events - which we refer to as Black Swan events - must necessarily occur. It follows as a corollary of our theorem that any computational model of decision-making under uncertainty is incomplete in the sense that not all events that occur can be taken into account. In the context of decision theory we argue that this constitutes a stronger sense of uncertainty than Knightian uncertainty.
\end{abstract}


\end{frontmatter}

\section{Introduction}

Ontology, loosely defined as ``\textit{the study of what there is}'' \citep{StanfordOntology}, studies questions of the existence of entities, their properties, and the relation between the two \citep{StanfordOntology}. So-called \textit{ontological arguments} traditionally are proofs of the existence of god, deducing this conclusion from a set of properties attributed to the entity `god'\footnote{See St. St. Anselm's proof \citep{Canterbury} for the most well-known example, or \cite{Goedel} for a recent presentation of such a proof originally devised by Kurt G\"{o}del.}. In this article we apply the same technique to show the existence of so-called \textbf{Black Swan} events.

The notion of Black Swan events, originally introduced in \cite{BlackSwan2005}, has been popularized by a series of popular science books \citep{Incerto}. Its formalization is work in progress by \cite{SilentRisk}. Their naming makes reference to the so-called \textbf{problem of induction}, defined by the Oxford English Dictionary as ``\textit{the process of inferring a general law or principle from the observation of particular instances}'' \citep{OED}. It is attributed to David Hume, who stated that such arguments cannot be made rigorous by deductive reasoning alone, given the lack for a justification of the assumption that yet unobserved entities will share the same properties as those already observed \citep{Hume}. The solution to the problem that this causes to scientific reasoning proposed by \cite{Popper} is to replace induction with falsification, the process of continuously trying to find empirical evidence against the theses of a scientific theory \citep{StanfordInduction}. This method was later rejected by critics as relying on a ``\textit{whiff}'' (some extent) of inductivism itself \citep{NewtonSmith1981,Salmon1981}.

The need for induction as part of the process as scientific discovery was already discussed by Aristotle, who argued that scientists should infer explanatory principles from phenomena in order to deduce further statements about them. Aristotle's method was generally accepted by medieval thinkers and many versions of such methods were presented by philosophers including Roger Bacon, Duns Scotus and William of Ockham, amongst others. The difficulty of arriving at general truths instead of accidental generalizations was generally understood by these authors \citep{Losee2001}. A common way of presenting this problem is to point out that the fact that all swans observed hitherto (in Europe) were white could lead an observer to induce that all swans are white -- a case of accidental predication that Duns Scotus seeked to avoid by stating that the most that could be inferred from observations was their ``\textit{aptitudinal union}'', in this case that swans \textit{could} be white \citep{Losee2001}. In this sense, the discovery of a species of black swans in Australia -- \textit{cygnus atratus} -- during the voyages of the 17th century is a good exemplification of this problem. First accounts of such sightings by the Dutch skipper Antony Caen in 1636 were met with skepticism back home, until Willem de Vlamingh brought back real specimen to the continent over 60 years later \citep{cygnus}. \cite{BlackSwan2005} states this historical context as a reason for choosing \textit{cygnus atratus} as the namesake for events with large impact, incomputable probabilities, and surprise effect properties.

The theory of Black Swan events, under development by \cite{SilentRisk}, has already made significant impact in popular language and general media. In this article we undertake to demonstrate that the occurrence of such events is in fact implied by their definition. Several authors have written about Black Swan events from a statistical and risk (management) theory perspective (apart from the already cited works, further examples include e.g. \cite{nafday2009strategies,NAFDAY2011108,HILAL20112374,Taleb2012,Aven2013}). \cite{BlackSwan2005,Yudkowsky2008} discuss Black Swans in the context of human cognition and its limitations. However, there are -- to the best of our knowledge -- no works looking at Black Swans from an analytical perspective.


\section{Definitions} \label{sec:Definitions}

A Black Swan is defined by \cite{SilentRisk} as ``\textit{a) a highly unexpected event} [f]\textit{or a given observer} [that] \textit{b) carries large consequences, and c) is subjected to ex-post rationalization}''.

In order to formalize this definition, we define a set $X$ of all events. We define the predicate $\chi (x)$ to denote that the event $x \in X$ can be \textbf{imagined} (by a given observer). We will discuss further below what it means in the context of our model to be able to imagine an event. In order to discuss the occurrence or non-occurrence of events we further define $\varphi (x)$ to denote an event that \textbf{occurs}. We make very few assumptions about $X$, which are summarized by the axioms presented in section \ref{sec:Reasoning}. 
Note that we do not include a notion of time in our theory, i.e. the chronological ordering of the occurrence of an event and its imagination by an observer carries no importance. We associate the non-imaginability of an event to the property of being highly unexpected (denoted (a) in the definition by \cite{SilentRisk}). We will see later that the property of large consequences (b) follows from this definition. We do not include ex-post rationalization (c) in our definition, as we do not consider it essential to the ontological nature of Black Swans. If we denote by $B(x)$ the property that event $x$ is a Black Swan, then its definition in our theory reads:

\[
B(x) \Leftrightarrow \neg \chi(x)
\]

In order to be able to discuss the `size' of consequences we introduce the partial order $<$ which satisfies the following axioms for all elements $a,b,c \in X$:

\begin{itemize}
\item Irreflexivity: $\neg (a < a)$
\item Transitivity: $(a < b) \land (b < c) \Rightarrow a < c$
\end{itemize}

Other than assuming a strict rather than a weak partial order, which we do for technical convenience here, this relation is consistent with von-Neumann-Morgenstern utility theory \citep{Neumann}. It shares the important transitivity property, a consistent equivalent for continuity under a strict partial order can be formulated, completeness and independence are not required but fully consistent with our theory. In this spirit, we will sometimes treat the notion of event $y$ having greater consequences than event $x$, i.e. $x<y$, as semantically equivalent $y$ being `worse' (i.e. yielding lower utility for a given observer) than $x$. This semantic interpretation does not affect the generality of the argument, and we stress that our argument does not require that the size of an event is in any way related to its utility for an observer. This interpretation does, however, help to emphasize the particular importance of unimaginable events when they are associated with negative outcomes (for a given observer). 

Our definition of Black Swan events may be seen as stricter than that of \cite{SilentRisk} in one sense, as non-imaginability can be seen as a stronger requirement than being highly unexpected. It may also be seen as wider in the sense that it does not require ex-post rationalization. In any case, we consider it a highly important class of events, as becomes clear when it is viewed in the context of decision-making under risk and uncertainty.

So-called \textbf{Knightian uncertainty} refers to the non-quantifiability of phenomena under conditions of uncertainty \citep{Knight1921}. Uncertainty in decision theory is often interpreted in the sense that the probability distribution over future events is unknown (i.e. not allowing for probabilistic calculations that would be possible under conditions of risk), while the set of possible future states and their respective payoffs are still known. This allows for the application of non-probabilistic computation models such as Wald's maximin-criterion or similar techniques \citep{Wald1939,Wald1945}. Black Swan events, as we consider them here, require a stronger sense of uncertainty, whereunder not even the full set of potential events or their payoffs are available to the decision maker. This notion may be seen as closer to the original definition of uncertainty by \cite{Knight1921}, which states that ``\textit{We} [...] \textit{restrict the term `uncertainty' to cases of the non-quantitive type.}''. It should be noted, however, that the emphasis here lies on a different, arguably even stronger point: the crux of Black Swan events, as defined herein, is not their non-quantifiability, but the fact that they cannot be considered in the decision-making process, regardless of whether quantitative or any other methods are used. \textbf{What we show in this paper is that there exist events which fundamentally cannot be taken into account when making decisions and which occur nonetheless}.

In order to formalize this idea, we define a standard computational model of decision-making comprising the following elements:

\begin{itemize}
\item A set $A$ of actions which are available to the agent. 
\item A set $P$ of information associated to the events (such as probabilities, or the property of occurrence).
\item A utility map $\Gamma$ which maps every pair of events and actions an outcome for a given agent: $\Gamma\colon (A,X) \rightarrow O$. Note that if we were to take the axioms of \cite{Neumann}, $\Gamma$ could be made consistent with a non-strict version of the size-relation $<$ introduced before, as shown in their proof. This is, however, not required for the point that we wish to make.
\item A decision map $\Phi$ which maps a vector of outcomes and a vector of associated information to the set of actions: $\Phi\colon (O^n,P^n) \rightarrow A$, where $n$ is the cardinality of the Cartesian product $A \times X$. In accordance with the concept of uncertainty defined above we may assume -- without loss of generality -- that there is no variation in the set of associated information $P$ and write $\Phi(O^n)$ for notational convenience.
\end{itemize}

We say that an event being non-imaginable for an agent is equivalent to her not being able to map it to an action
. 
We consider this as being different from deciding not to react to an event, because the latter entails finishing the computation $\Gamma$ of an outcome, which can then be mapped to whatever action would have been chosen without knowledge of the event (one may also think of $A$ containing another response labeled `Do Nothing'). We make this distinction because it facilitates the discussion of the computational aspects of decision-making, which we present in section \ref{sec:Horatio}, but note that it is not essential to the validity of our argument. For now we present the definition of the decision map, which states that a given set of events has to contain at least one imaginable event in order for it to be mapped to an action:

\[
\Phi(\Gamma^n(A,X)) = \begin{cases} \uparrow & \mbox{if } \forall x \in X \colon \neg \chi(x) \\ a \in A & \mbox{otherwise} \end{cases},
\]

where $\Gamma^n(A,X)$ denotes the element-wise application of the $\Gamma$ map to every element in the Cartesian product $A \times X$, and $\uparrow$ denotes the fact that the computation has not terminated.

\section{Reasoning} \label{sec:Reasoning}






Our reasoning can be compactly summarized as follows:

\textbf{Axiom 1}: \\There exists (at least) one event that occurs that is so bad that an event with greater consequences cannot be imagined:

\[
\exists x (\varphi(x) \land \forall y (x<y \rightarrow \neg \chi(y)))
\]

\textbf{Axiom 2}: \\No matter how bad an event that occurs is, there exists an event with even greater consequences which occurs:

\[
\forall x (\varphi(x) \rightarrow \exists y (\varphi(y) \land x < y))
\]

\textbf{Theorem}: \\
Black Swan events occur: 

\[
\exists x (B(x) \land \varphi(x))
\]

In \ref{sec:Proof}, we present a formal proof of the above argument using a Hilbert-style system. Here we give a proof via semantic argument. 

\begin{proof} \leavevmode

By axiom 1, there exists an element $x \in X$ such that: 
\begin{itemize}
\item[1.] $\varphi(x)$
\item[2.] $\forall y (x<y \rightarrow \neg \chi(y))$.
\end{itemize}
By 1. and axiom 2 we obtain that there exists $y \in X$ such that $\varphi(y)$ and $x<y$. By the latter, and by 2., we obtain $\neg \chi(y)$. Therefore we have obtained $y \in X$ such that $\neg \chi(y) \land \varphi(y)$, that is $B(y) \land \varphi(y)$. This shows that axioms 1 and 2 imply our Theorem. 
\end{proof}

We now move on to lay out the rationale underpinning our axioms.

\subsection{Axiom 1} \label{sec:Horatio}

Axiom 1 states that all events greater than a particular event (which occurs) are beyond our imagination, regardless of whether they occur or not. We call this argument the \textbf{Horatio-Principle}, after the following quote:

\begin{centering}
\textit{``There are more things in heaven and earth, Horatio, than are dreamt of in your philosophy.''}\\
\end{centering}
\hspace{12cm}{- \textbf{Hamlet (1.5.167-8)}}

It has been argued that Hamlet in this quote is talking about general limitations of human thought rather than trying to insult Horatio's intellect \citep{HamletBradley}, and some versions of the text even talk about `our' instead of `your' philosophy\footnote{The Folio version of Hamlet uses `our', while the first and second Quarto versions use `your'. While it is unclear whether this difference stems from an editor's mistake, it is commonly understood that the usually adopted `your' is meant as a general address and not and not as a direct attack on Horatio \citep{HamletThompson}.}. The Horatio-Principle here states the existence of events that are `not dreamt of' in the `philosophy' of a given observer, i.e. which she cannot imagine.

In order to justify the Horatio-Principle, we refer to the computation model described by the decision map $\Phi$ introduced in section \ref{sec:Definitions}. As stated in the definition, the notion of being able to imagine an event is equivalent to being able to come up with a response to that event. The decision making process could thus be viewed as a Turing machine \citep{Turing1937} that tries to compute a response to a given set of events. In this case $A$ would be the set of terminating states of the Turing machine, $X$ would be the set of alphabet symbols, which here also serves as the set of input symbols, and $\Phi$ the transition function. As stated by the Halting problem \citep{Turing1937}, no such machine could be guaranteed to ever reach a terminating state, i.e. to arrive at a decision upon a given set of events. The Horatio principle states (i) that there exist events for which a response is not computable, which is motivated by the Halting problem. Furthermore (ii), it states that there exists a size threshold for the consequences of an event beyond which this holds for all events (i.e. that the computability of an event is to some extent proportional to the size of its outcome). And lastly (iii), it states that there exists an event which does not exceed the size threshold and which has the property of occurring. Hence, while the Halting problem does not fully extend to the Horatio principle, we consider it a strong motivation thereof.

One may be lead to think that the possibility of deliberate \textbf{Antifragility} \citep{Incerto} might induce a type of Russell paradox \citep{Russell} in our system: we can imagine the occurrence of a Black Swan event and thus adjust our strategies accordingly. Antifragile strategies allow us to benefit from such an occurrence, even though we would not be able to describe the nature of the event in advance. Our set of imaginable events includes the occurrence of events that are not imaginable. Imagination in this case does not specify the event itself, which would thus still be outside the set of imaginable events. According to the Horatio-Principle, this set is still non-empty. In other words, even for (seemingly) antifragile strategies, there exist events with large consequences for which the outcome cannot be known before they occur
.

\subsection{Axiom 2}

Axiom 2 states that for every event that occurs, an event that is greater occurs as well. We argue that the set of occurring events is an open set and call this assumption \textbf{Murphy's law}, the often humorously stated aphorism that ``\textit{anything that can go wrong will go wrong}''. Formally, Axiom 2 can be derived from Murphy's law, as expressed in our system, by using an additional assumption that set of events $X$ is an open set (the formal proof is left to the reader):

\begin{itemize}
\item $\forall x,y (\phi(x) \land (x < y) \supset \phi(y))$ \\ Murphy's law
\item $\forall x \exists y (x < y)$ \\ Open universe
\item $(\forall x) \Big( \varphi (x) \supset (\exists y) (\varphi (y) \land (x<y)) \Big)$ \\ Axiom 2
\end{itemize}



\section{Implications}

In this section we show that it follows as a corollary of our Theorem that every computational model of decision-making is incomplete in the sense that not all events can be taken into account, and that this also concerns events that do occur.

A computational model is said to be \textbf{complete} if the decision map $\Phi$ has the property that for every two sets of events which differ by at least one element $Y,Z\colon (\exists y \in Y\colon y \notin Z) \lor (\exists z \in Z\colon z \notin Y)$ there exist a set of actions $A$ and a map $\Gamma$ such that $\Phi(\Gamma^n(A,Y)) \ne \Phi(\Gamma^n(A,Z))$. This means that a complete decision-model always allows an agent to act differently under different circumstances if this is indicated by her preferences, as expressed by the utility map $\Gamma$. In other words, a complete decision map $\Phi(\Gamma^n(A,X))$ takes into account all events in $X$. A more refined concept of \textbf{completeness with respect to occurring events} only requires this for events that have the property of occurring, which means that for $Y,Z$ such that $\forall y \in Y : \varphi(y)$ and $ \forall z \in Z : \varphi(z) $, we have $\Phi(\Gamma^n(A,Y)) \ne \Phi(\Gamma^n(A,Z))$ if and only if $Y \neq Z$.

Assume for the sake of contradiction that there exists a computational model with a complete decision map $\Phi\colon (O^n,P^n) \rightarrow A$, as defined above. Let $S\subseteq X$ be a set of events which occur and which are not imaginable (i.e. Black Swan events) for a given agent: $S = \{s\colon s \in X \land \varphi(s) \land \neg\chi(s)\}$. It follows from our theorem that the set $S$ is non-empty. It further follows from Axiom 2 that for every event in $S$ there exists an event with greater consequences that occurs as well. The cardinality of $S$ is thus at least $\aleph_0$. Thus there exist two vectors of unimaginable events which differ by at least one element $Y,Z \in S^j \colon (\exists y \in Y\colon y \notin Z) \lor (\exists z \in Z\colon z \notin Y)$. By the definition of $\Phi$, both vectors are not mapped to any response, because the computation will not terminate. Therefore it is impossible to have $\Phi(\Gamma^n(A,Y)) \ne \Phi(\Gamma^n(A,Z))$, contradicting the assumption of the existence of a complete decision-making model with respect to occurring events. The contradiction of the existence of a complete decision-making model can be obtained by setting $S=\{s\colon s \in X \land \neg\chi(s)\}$.

\section{Conclusion}

We developed a first-order deductive system to show that the occurrence of Black Swan events is implied by their definition. We make two assumptions, namely that our imagination is bounded and that the universe of occurring events is an open set, which we call the Horatio Principle and Murphy's law, respectively. We motivate the Horatio principle by showing that under a computational model of human decision-making, the question of whether all events are imaginable can be reduced to the Halting problem. We present a formal proof of our argument using a Hilbert System. We show that it follows as a corollary of this Theorem that every computational model of decision-making is incomplete in the sense that not all events that occur can be taken into account in the decision-making process. When viewed through the lense of decision-making under uncertainty -- as in von-Neumann-Morgenstern utility theory -- we argue that Black Swans entail a stronger sense of uncertainty than Knightian uncertainty because their existence means that even under perfect information no decision criterion -- regardless of whether it is of quantitative nature or not -- can make use of all the information available.

\section*{Acknowledgments}

The idea for this paper developed out of a conversation with Davoud Taghawi-Nejad at the Institute for New Economic Thinking at the Oxford Martin School. We thank Prof. Timothy Williamson for his comments on the paper. We further thank Matthew Deakin and G\"unther Siebenbrunner for useful remarks that have been incorporated into the paper.

\section*{References}\label{sec_References}
\bibliographystyle{apalike}
\bibliography{Mendeley,literature}

\appendix
\section{Formal proof} \label{sec:Proof}

In this section we present the formal system which in which we will establish the theorem of the occurrence of Black Swan events and its proof.

\subsection{Hilbert System}

We use a deductive system for first-order logic. This system consists of the following axioms and rules.

\subsubsection{Axioms}

\begin{itemize}
    \item[(FO1)] $A \supset (B \supset A)$
    \item[(FO2)] $(A \supset B) \supset (A \supset (B \supset C)) \supset (A \supset C)$
    \item[(FO3)] $A \supset (A \lor B)$
    \item[(FO4)] $B \supset (A \lor B)$
    \item[(FO5)] $(A \supset C) \supset (B \supset C) \supset (A \lor B \supset C)$
    \item[(FO6)] $(A \supset B) \supset (A \supset \neg B) \supset \neg A$
    \item[(FO7)] $\neg \neg A \supset A$
    \item[(FO8)] $A \land B \supset A$
    \item[(FO9)] $A \land B \supset B$
    \item[(FO10)] $A \supset B \supset (A \land B)$
    \item[(FO11)] $A(t) \supset (\exists x)A(x)$ where $t$ can be any term
    \item[(FO12)] $(\forall x)A(a) \supset A(t)$ where $t$ can be any term
\end{itemize}

\subsubsection{Rules}

\begin{itemize}
    \item[(MP)] $\frac{A \quad A \supset B}{B}$
    \item[(R1)] $\frac{C \ \supset \ A(x)}{C \ \supset \ (\forall x)A(x)}$ where the variable $x$ must not occur free in $C$
    \item[(R2)] $\frac{A(x) \ \supset \ C}{(\exists x)A(x) \ \supset \ C}$ where the variable $x$ must not occur free in $C$
    \item[(R3)]$\frac{(A \land B) \supset C}{B \supset A \supset C}$
\end{itemize}

\subsection{Proof of the theorem}

\newcommand{\recurrent}{\varphi(x) \land (\forall y) ((x<y) \supset \neg \chi(y))}

Using the deductive system described above, we show a proof of our theorem
\[\mathrm{Thm} := (\exists z) (\varphi (z) \land \neg \chi (z)) \ . \]

Our \emph{proof} will be a list of formulas $A_1, \dots , A_n$ such that:
\begin{itemize}
    \item $A_n = \text{Thm}$
    \item for every $i=1 \dots n$, $A_i$ is either one of the axioms (FO1)--(FO12), or it is one of our two axioms
    \[ \mathrm{Ax1} := (\exists x) \Big(\recurrent \Big)  \] 
    \[ \mathrm{Ax2} := (\forall x) \Big( \varphi (x) \supset (\exists y) (\varphi (y) \land (x<y)) \Big) \ ,\]
    or it is deduced by applying one of the rules (MP),(R1)--(R3) to some formula(s) $A_1,\dots,A_{i-1}$.
\end{itemize}

\newcommand{\spazio}{\mathrm{ }}

\newcommand{\lungo}{\Big( \recurrent \Big) \land ( \varphi (y) \land (x<y) )}

\begin{proof}

$ \ $

\begin{itemize}
    \item[1.] $\mathrm{Ax2} \supset \Big( \varphi (x) \supset (\exists y) (\varphi (y) \land (x<y)) \Big)$ \\
    axiom (FO12)
    \item[2.] $\mathrm{Ax2}$ \\
    axiom (Ax2)
    \item[3.] $\varphi (x) \supset (\exists y) (\varphi (y) \land (x<y))$ \\
    by (MP) from 2. and 1.
    \item[4.] $\Big( \varphi (x) \supset (\exists y) (\varphi (y) \land (x<y)) \Big) \supset \Big[ \Big( \recurrent \Big) \supset \Big( \varphi (x) \supset (\exists y) (\varphi (y) \land (x<y)) \Big) \Big]$ \\
    axiom (FO1)
    \item[5.] $\Big( \recurrent \Big) \supset \Big( \varphi (x) \supset (\exists y) (\varphi (y) \land (x<y)) \Big)$ \\
    by (MP) from 3. and 4.
    \item[6.] $\Big( \recurrent \Big) \supset \varphi (x)$ \\
    axiom (FO8)
    \item[7.] $ \Big[ \Big( \recurrent \Big) \supset \varphi (x) \Big] \supset $ \\
    $\spazio$ $\spazio$ $\spazio$ $ \Big[ \Big( \recurrent \Big) \supset \Big( \varphi (x) \supset (\exists y) (\varphi (y) \land (x<y)) \Big) \Big] \supset $ \\
    $\spazio$ $\spazio$ $\spazio$ $\spazio$ $\spazio$ $\spazio$ $ \Big[ \Big( \recurrent \Big) \supset (\exists y) (\varphi (y) \land (x<y)) \Big] $ \\
    axiom (FO2)
    \item[8.] $ \Big[ \Big( \recurrent \Big) \supset \Big( \varphi (x) \supset (\exists y) (\varphi (y) \land (x<y)) \Big) \Big] \supset $ \\
    $\spazio$ $\spazio$ $\spazio$ $ \Big[ \Big( \recurrent \Big) \supset (\exists y) (\varphi (y) \land (x<y)) \Big] $ \\
    by (MP) from 6. and 7. 
    \item[9.] $ \Big( \recurrent \Big) \supset (\exists y) (\varphi (y) \land (x<y))$ \\
    by (MP) from 5. and 8.
    \item[10.] $(\varphi (y) \land (x<y)) \supset (x<y)$ \\
    axiom (FO9)
    \item[11.] $ \Big( \recurrent \Big) \supset (\forall y) ((x<y) \supset \neg \chi (y)) $ \\
    axiom (FO9)
    \item[12.] $(\forall y) ((x<y) \supset \neg \chi (y)) \supset ((x<y) \supset \neg \chi (y))$ \\
    axiom (FO12)
    \item[13.] $\Big[ (\forall y) ((x<y) \supset \neg \chi (y)) \supset ((x<y) \supset \neg \chi (y)) \Big] \supset$ \\ 
    $\spazio$ $\spazio$ $\spazio$ $\Big( \recurrent \Big) \supset \Big[ (\forall y) ((x<y) \supset \neg \chi (y)) \supset ((x<y) \supset \neg \chi (y)) \Big]$ \\
    axiom (FO1)
    \item[14.] $\Big( \recurrent \Big) \supset \Big[ (\forall y) ((x<y) \supset \neg \chi (y)) \supset ((x<y) \supset \neg \chi (y)) \Big]$ \\
    by (MP) from 12. and 13.
    \item[15.] $\Big[ \Big( \recurrent \Big) \supset (\forall y) ((x<y) \supset \neg \chi (y)) \Big] \supset$ \\
    $\spazio$ $\spazio$ $\spazio$ $ \Big[ \Big( \recurrent \Big) \supset \Big[ (\forall y) ((x<y) \supset \neg \chi (y)) \supset ((x<y) \supset \neg \chi (y)) \Big] \Big] \supset$ \\
    $\spazio$ $\spazio$ $\spazio$ $\spazio$ $\spazio$ $\spazio$ $\Big[ \Big( \recurrent \Big) \supset ((x<y) \supset \neg \chi (y)) \Big]$ \\
    axiom (FO2)
    \item[16.] $\Big[ \Big( \recurrent \Big) \supset \Big[ (\forall y) ((x<y) \supset \neg \chi (y)) \supset ((x<y) \supset \neg \chi (y)) \Big] \Big] \supset$ \\
    $\spazio$ $\spazio$ $\spazio$ $\Big[ \Big( \recurrent \Big) \supset ((x<y) \supset \neg \chi (y)) \Big]$ \\
    by (MP) from 11. and 15.
    \item[17.] $ \Big( \recurrent \Big) \supset ((x<y) \supset \neg \chi (y)) $ \\
    by (MP) from 14. and 16.
    \item[18.] $\Big( (\varphi (y) \land (x<y)) \supset (x<y) \Big) \supset $ \\
    $\spazio$ $\spazio$ $\spazio$ $\Big[ \Big( \recurrent \Big) \land (\varphi (y) \land (x<y)) \Big] \supset \Big( (\varphi (y) \land (x<y)) \supset (x<y) \Big)$ \\
    axiom (FO1)
    \item[19.] $\Big[ \Big( \recurrent \Big) \land (\varphi (y) \land (x<y)) \Big] \supset \Big( (\varphi (y) \land (x<y)) \supset (x<y) \Big)$ \\
    by (MP) from 10. and 18.
    \item[20.] $\Big[ \Big( \recurrent \Big) \land (\varphi (y) \land (x<y)) \Big] \supset (\varphi (y) \land (x<y))$ \\
    axiom (FO9)
    \item[21.] $\Big[ \Big[ \Big( \recurrent \Big) \land (\varphi (y) \land (x<y)) \Big] \supset (\varphi (y) \land (x<y)) \Big] \supset$ \\
    $\spazio$ $\spazio$ $\spazio$ $\Big[ \Big[ \Big( \recurrent \Big) \land (\varphi (y) \land (x<y)) \Big] \supset \Big( (\varphi (y) \land (x<y)) \supset (x<y) \Big)  \Big] \supset$ \\
    $\spazio$ $\spazio$ $\spazio$ $\spazio$ $\spazio$ $\spazio$ $\Big[ \Big( \recurrent \Big) \land (\varphi (y) \land (x<y)) \Big] \supset (x<y)$ \\
    axiom (FO2)
    \item[22.] $\Big[ \Big[ \Big( \recurrent \Big) \land (\varphi (y) \land (x<y)) \Big] \supset \Big( (\varphi (y) \land (x<y)) \supset (x<y) \Big)  \Big] \supset$ \\
    $\spazio$ $\spazio$ $\spazio$ $\Big[ \Big( \recurrent \Big) \land (\varphi (y) \land (x<y)) \Big] \supset (x<y)$ \\
    by (MP) from 20. and 21.
    \item[23.] $\Big[ \Big( \recurrent \Big) \land (\varphi (y) \land (x<y)) \Big] \supset (x<y)$ \\
    by (MP) from 19. and 22.
    \item[24.] $\Big[ \Big( \recurrent \Big) \supset ((x<y) \supset \neg \chi (y)) \Big] \supset$ \\
    $\spazio$ $\spazio$ $\spazio$ $\Big[ \Big( \recurrent \Big) \land (\varphi (y) \land (x<y)) \Big] \supset$ \\
    $\spazio$ $\spazio$ $\spazio$ $\spazio$ $\spazio$ $\spazio$ $\Big[ \Big( \recurrent \Big) \supset ((x<y) \supset \neg \chi (y)) \Big]$ \\
    axiom (FO1)
    \item[25.] $\Big[ \Big( \recurrent \Big) \land (\varphi (y) \land (x<y)) \Big] \supset$ \\
    $\spazio$ $\spazio$ $\spazio$ $\Big[ \Big( \recurrent \Big) \supset ((x<y) \supset \neg \chi (y)) \Big]$ \\
    by (MP) from 17. and 24.
    \item[26.] $\Big[ \Big( \recurrent \Big) \land (\varphi (y) \land (x<y)) \Big] \supset \Big( \recurrent \Big) $ \\
    axiom (FO8)
    \item[27.] $\Big[ \Big[ \Big( \recurrent \Big) \land (\varphi (y) \land (x<y)) \Big] \supset \Big( \recurrent \Big) \Big] \supset$ \\
    $\spazio$ $\spazio$ $\spazio$ $\Bigg[ \Big[ \Big( \recurrent \Big) \land (\varphi (y) \land (x<y)) \Big] \supset$ \\
    $\spazio$ $\spazio$ $\spazio$ $\spazio$ $\spazio$ $\spazio$ $\Big[ \Big( \recurrent \Big) \supset ((x<y) \supset \neg \chi (y)) \Big] \Bigg] \supset$ \\
    $\spazio$ $\spazio$ $\spazio$ $\spazio$ $\spazio$ $\spazio$ $\spazio$ $\spazio$ $\spazio$ $\Big[ \Big[ \Big( \recurrent \Big) \land (\varphi (y) \land (x<y)) \Big] \supset ((x<y) \supset \neg \chi (y)) \Big]$ \\
    axiom (FO2)
    \item[28.] $\Bigg[ \Big[ \Big( \recurrent \Big) \land (\varphi (y) \land (x<y)) \Big] \supset$ \\
    $\spazio$ $\spazio$ $\spazio$ $\Big[ \Big( \recurrent \Big) \supset ((x<y) \supset \neg \chi (y)) \Big] \Bigg] \supset$ \\
    $\spazio$ $\spazio$ $\spazio$ $\spazio$ $\spazio$ $\spazio$ $\Big[ \Big[ \Big( \recurrent \Big) \land (\varphi (y) \land (x<y)) \Big] \supset ((x<y) \supset \neg \chi (y)) \Big]$ \\
    by (MP) from 26. and 27.
    \item[29.] $\Big[ \Big( \recurrent \Big) \land (\varphi (y) \land (x<y)) \Big] \supset ((x<y) \supset \neg \chi (y))$ \\
    by (MP) from 25. and 28.
    \item[30.] $\Big[ \Big[ \Big( \recurrent \Big) \land (\varphi (y) \land (x<y)) \Big] \supset (x<y) \Big] \supset$ \\
    $\spazio$ $\spazio$ $\spazio$ $\Big[ \Big[ \Big( \recurrent \Big) \land (\varphi (y) \land (x<y)) \Big] \supset ((x<y) \supset \neg \chi (y)) \Big] \supset$ \\
    $\spazio$ $\spazio$ $\spazio$ $\spazio$ $\spazio$ $\spazio$ $\Big[ \Big( \recurrent \Big) \land (\varphi (y) \land (x<y)) \Big] \supset \neg \chi (y)$ \\
    axiom (FO2)
    \item[31.] $\Big[ \Big[ \Big( \recurrent \Big) \land (\varphi (y) \land (x<y)) \Big] \supset ((x<y) \supset \neg \chi (y)) \Big] \supset$ \\
    $\spazio$ $\spazio$ $\spazio$ $\Big[ \Big( \recurrent \Big) \land (\varphi (y) \land (x<y)) \Big] \supset \neg \chi (y)$ \\
    by (MP) from 23. and 30.
    \item[32.] $\Big[ \lungo \Big] \supset \neg \chi (y)$ \\
    by (MP) from 29. and 31.
    \item[33.] $(\varphi (y) \land (x<y)) \supset \varphi (y)$ \\
    axiom (FO8)
    \item[34.] $\Big( (\varphi (y) \land (x<y)) \supset \varphi (y) \Big) \supset$ \\
    $\spazio$ $\spazio$ $\spazio$ $\Big[ \Big( \recurrent \Big) \land (\varphi (y) \land (x<y)) \Big] \supset \Big( (\varphi (y) \land (x<y)) \supset \varphi (y) \Big)$ \\
    axiom (FO1)
    \item[35.] $\Big[ \Big( \recurrent \Big) \land (\varphi (y) \land (x<y)) \Big] \supset \Big( (\varphi (y) \land (x<y)) \supset \varphi (y) \Big)$ \\
    by (MP) from 33. and 34.
    \item[36.] $\Big[ \Big( \recurrent \Big) \land (\varphi (y) \land (x<y)) \Big] \supset (\varphi (y) \land (x<y))$ \\
    axiom (FO9)
    \item[37.] $\Bigg[ \Big[ \Big( \recurrent \Big) \land (\varphi (y) \land (x<y)) \Big] \supset (\varphi (y) \land (x<y)) \Bigg] \supset$ \\
    $\spazio$ $\spazio$ $\spazio$ $\Bigg[ \Big[ \Big( \recurrent \Big) \land (\varphi (y) \land (x<y)) \Big] \supset \Big( (\varphi (y) \land (x<y)) \supset \varphi (y) \Big) \Bigg] \supset$ \\
    $\spazio$ $\spazio$ $\spazio$ $\spazio$ $\spazio$ $\spazio$ $\Bigg[ \Big[ \Big( \recurrent \Big) \land (\varphi (y) \land (x<y)) \Big] \supset \varphi (y) \Bigg]$ \\
    axiom (FO2)
    \item[38.] $\Bigg[ \Big[ \Big( \recurrent \Big) \land (\varphi (y) \land (x<y)) \Big] \supset \Big( (\varphi (y) \land (x<y)) \supset \varphi (y) \Big) \Bigg] \supset$ \\
    $\spazio$ $\spazio$ $\spazio$ $\Bigg[ \Big[ \Big( \recurrent \Big) \land (\varphi (y) \land (x<y)) \Big] \supset \varphi (y) \Bigg]$ \\
    by (MP) from 36. and 37.
    \item[39.] $\Big[ \Big( \recurrent \Big) \land (\varphi (y) \land (x<y)) \Big] \supset \varphi (y)$ \\
    by (MP) from 35. and 38.
    \item[40.] $\varphi (y) \supset \neg \chi (y) \supset ( \varphi (y) \land \neg \chi (y) )$ \\
    axiom (FO10)
    \item[41.] $\Big( \varphi (y) \supset \neg \chi (y) \supset ( \varphi (y) \land \neg \chi (y) ) \Big) \supset$ \\
    $\spazio$ $\spazio$ $\spazio$ $\Big[ \lungo \Big] \supset \Big( \varphi (y) \supset \neg \chi (y) \supset ( \varphi (y) \land \neg \chi (y) ) \Big)$ \\
    axiom (FO1)
    \item[42.] $\Big[ \lungo \Big] \supset \Big( \varphi (y) \supset \neg \chi (y) \supset ( \varphi (y) \land \neg \chi (y) ) \Big)$ \\
    by (MP) from 40. and 41. 
    \item[43.] $\Bigg[ \Big[ \Big( \recurrent \Big) \land (\varphi (y) \land (x<y)) \Big] \supset \varphi (y) \Bigg] \supset$ \\
    $\spazio$ $\spazio$ $\spazio$ $\Bigg[ \Big[ \lungo \Big] \supset \Big( \varphi (y) \supset \neg \chi (y) \supset ( \varphi (y) \land \neg \chi (y) ) \Big) \Bigg] \supset$ \\
    $\spazio$ $\spazio$ $\spazio$ $\spazio$ $\spazio$ $\spazio$ $\Bigg[ \Big[ \lungo \Big] \supset \Big( \neg \chi (y) \supset ( \varphi (y) \land \neg \chi (y) ) \Big) \Bigg]$ \\
    axiom (FO2)
    \item[44.] $\Bigg[ \Big[ \lungo \Big] \supset \Big( \varphi (y) \supset \neg \chi (y) \supset ( \varphi (y) \land \neg \chi (y) ) \Big) \Bigg] \supset$ \\
    $\spazio$ $\spazio$ $\spazio$ $\Bigg[ \Big[ \lungo \Big] \supset \Big( \neg \chi (y) \supset ( \varphi (y) \land \neg \chi (y) ) \Big) \Bigg]$ \\
    by (MP) from 39. and 43.
    \item[45.] $\Big[ \lungo \Big] \supset \Big( \neg \chi (y) \supset ( \varphi (y) \land \neg \chi (y) ) \Big)$ \\
    by (MP) from 42. and 44.
    \item[46.] $\Bigg[ \Big[ \lungo \Big] \supset \neg \chi (y) \Bigg] \supset$ \\
    $\spazio$ $\spazio$ $\spazio$ $\Bigg[ \Big[ \lungo \Big] \supset \Big( \neg \chi (y) \supset ( \varphi (y) \land \neg \chi (y) ) \Big) \Bigg]$ \\
    $\spazio$ $\spazio$ $\spazio$ $\spazio$ $\spazio$ $\spazio$ $\Bigg[ \Big[ \lungo \Big] \supset (\varphi (y) \land \neg \chi (y)) \Bigg]$ \\
    axiom (FO2)
    \item[47.] $\Bigg[ \Big[ \lungo \Big] \supset \Big( \neg \chi (y) \supset ( \varphi (y) \land \neg \chi (y) ) \Big) \Bigg]$ \\
    $\spazio$ $\spazio$ $\spazio$ $\Bigg[ \Big[ \lungo \Big] \supset (\varphi (y) \land \neg \chi (y)) \Bigg]$ \\
    by (MP) from 32. and 46.
    \item[48.] $\Big[ \lungo \Big] \supset (\varphi (y) \land \neg \chi (y))$ \\
    by (MP) from 45. and 47.
    \item[49.] $(\varphi (y) \land \neg \chi (y)) \supset (\exists z) (\varphi (z) \land \neg \chi (z))$\\
    axiom (FO11)
    \item[50.] $\Big( (\varphi (y) \land \neg \chi (y)) \supset (\exists z) (\varphi (z) \land \neg \chi (z)) \Big) \supset$\\
    $\spazio$ $\spazio$ $\spazio$ $\Big[ \lungo \Big] \supset \Big( (\varphi (y) \land \neg \chi (y)) \supset (\exists z) (\varphi (z) \land \neg \chi (z)) \Big)$ \\
    axiom (FO1)
    \item[51.] $\Big[ \lungo \Big] \supset \Big( (\varphi (y) \land \neg \chi (y)) \supset (\exists z) (\varphi (z) \land \neg \chi (z)) \Big)$ \\
    by (MP) from 49. and 50.
    \item[52.] $\Bigg[ \Big[ \lungo \Big] \supset (\varphi (y) \land \neg \chi (y)) \Bigg] \supset$ \\
    $\spazio$ $\spazio$ $\spazio$ $\Bigg[ \Big[ \lungo \Big] \supset \Big( (\varphi (y) \land \neg \chi (y)) \supset (\exists z) (\varphi (z) \land \neg \chi (z)) \Big) \Bigg] \supset$ \\
    $\spazio$ $\spazio$ $\spazio$ $\spazio$ $\spazio$ $\spazio$ $\Bigg[ \Big[ \lungo \Big] \supset (\exists z) (\varphi(z) \land \neg \chi(z)) \Bigg]$ \\
    axiom (FO2)
    \item[53.] $\Bigg[ \Big[ \lungo \Big] \supset \Big( (\varphi (y) \land \neg \chi (y)) \supset (\exists z) (\varphi (z) \land \neg \chi (z)) \Big) \Bigg] \supset$ \\
    $\spazio$ $\spazio$ $\spazio$ $\Bigg[ \Big[ \lungo \Big] \supset (\exists z) (\varphi (z) \land \neg \chi (z)) \Bigg]$ \\
    by (MP) from 48. and 52.
    \item[54.] $\Big[ \lungo \Big] \supset (\exists z) (\varphi (z) \land \neg \chi (z))$ \\
    by (MP) from 51. and 53.
    \item[55.] $(\varphi (y) \land (x<y)) \supset \Big( \recurrent \Big) \supset (\exists z) (\varphi (z) \land \neg \chi (z))$ \\
    by (R3) from 54.
    \item[56.] $(\exists y) (\varphi (y) \land (x<y)) \supset \Big( \recurrent \Big) \supset (\exists z) (\varphi (z) \land \neg \chi (z))$ \\
    by (R2) from 55.
    \item[57.] $\Big[ (\exists y) (\varphi (y) \land (x<y)) \land \Big( \recurrent \Big) \Big] \supset  (\exists y) (\varphi (y) \land (x<y))$ \\
    axiom (FO8)
    \item[58.] $\Big[ (\exists y) (\varphi (y) \land (x<y)) \supset \Big( \recurrent \Big) \supset (\exists z) (\varphi (z) \land \neg \chi (z)) \Big] \supset$ \\
    $\spazio$ $\spazio$ $\spazio$ $\Big[ (\exists y) (\varphi (y) \land (x<y)) \land \Big( \recurrent \Big) \Big] \supset$ \\
    $\spazio$ $\spazio$ $\spazio$ $\spazio$ $\spazio$ $\spazio$ $\Big[ (\exists y) (\varphi (y) \land (x<y)) \supset \Big( \recurrent \Big) \supset (\exists z) (\varphi (z) \land \neg \chi (z)) \Big]$ \\
    axiom (FO1)
    \item[59.] $\Big[ (\exists y) (\varphi (y) \land (x<y)) \land \Big( \recurrent \Big) \Big] \supset$ \\
    $\spazio$ $\spazio$ $\spazio$ $\Big[ (\exists y) (\varphi (y) \land (x<y)) \supset \Big( \recurrent \Big) \supset (\exists z) (\varphi (z) \land \neg \chi (z)) \Big]$ \\
    by (MP) from 56. and 58.
    \item[60.] $\Bigg[ \Big[ (\exists y) (\varphi (y) \land (x<y)) \land \Big( \recurrent \Big) \Big] \supset  (\exists y) (\varphi (y) \land (x<y)) \Bigg] \supset$ \\
    $\spazio$ $\spazio$ $\spazio$ $\Bigg[ \Big[ (\exists y) (\varphi (y) \land (x<y)) \land \Big( \recurrent \Big) \Big] \supset$ \\
    $\spazio$ $\spazio$ $\spazio$ $\spazio$ $\spazio$ $\spazio$ $\Big[ (\exists y) (\varphi (y) \land (x<y)) \supset \Big( \recurrent \Big) \supset (\exists z) (\varphi (z) \land \neg \chi (z)) \Big] \Bigg] \supset$ \\
    $\spazio$ $\spazio$ $\spazio$ $\spazio$ $\spazio$ $\spazio$ $\Bigg[ \Big[ (\exists y) (\varphi (y) \land (x<y)) \land \Big( \recurrent \Big) \Big] \supset$ \\
    $\spazio$ $\spazio$ $\spazio$ $\spazio$ $\spazio$ $\spazio$ $\spazio$ $\spazio$ $\spazio$ $ \Big( \recurrent \Big) \supset (\exists z) (\varphi (z) \land \neg \chi (z)) \Bigg]$ \\
    axiom (FO2)
    \item[61.] $\Bigg[ \Big[ (\exists y) (\varphi (y) \land (x<y)) \land \Big( \recurrent \Big) \Big] \supset$ \\
    $\spazio$ $\spazio$ $\spazio$ $\Big[ (\exists y) (\varphi (y) \land (x<y)) \supset \Big( \recurrent \Big) \supset (\exists z) (\varphi (z) \land \neg \chi (z)) \Big] \Bigg] \supset$ \\
    $\spazio$ $\spazio$ $\spazio$ $\Bigg[ \Big[ (\exists y) (\varphi (y) \land (x<y)) \land \Big( \recurrent \Big) \Big] \supset$ \\
    $\spazio$ $\spazio$ $\spazio$ $\spazio$ $\spazio$ $\spazio$ $ \Big( \recurrent \Big) \supset (\exists z) (\varphi (z) \land \neg \chi (z)) \Bigg]$ \\
    by (MP) from 57. and 60.
    \item[62.] $\Big[ (\exists y) (\varphi (y) \land (x<y)) \land \Big( \recurrent \Big) \Big] \supset$ \\
    $\spazio$ $\spazio$ $\spazio$ $ \Big( \recurrent \Big) \supset (\exists z) (\varphi (z) \land \neg \chi (z))$ \\
    by (MP) from 59. and 61.
    \item[63.] $\Big[ (\exists y) (\varphi (y) \land (x<y)) \land \Big( \recurrent \Big) \Big] \supset \Big( \recurrent \Big)$ \\
    axiom (FO9)
    \item[64.] $\Bigg[ \Big[ (\exists y) (\varphi (y) \land (x<y)) \land \Big( \recurrent \Big) \Big] \supset \Big( \recurrent \Big) \Bigg] \supset$ \\
    $\spazio$ $\spazio$ $\spazio$ $\Bigg[ \Big[ (\exists y) (\varphi (y) \land (x<y)) \land \Big( \recurrent \Big) \Big] \supset$ \\  
    $\spazio$ $\spazio$ $\spazio$ $\spazio$ $\spazio$ $\spazio$ $ \Big( \recurrent \Big) \supset (\exists z) (\varphi (z) \land \neg \chi (z)) \Bigg] \supset$ \\
    $\spazio$ $\spazio$ $\spazio$ $\spazio$ $\spazio$ $\spazio$ $\Bigg[ \Big[ (\exists y) (\varphi (y) \land (x<y)) \land \Big( \recurrent \Big) \Big] \supset (\exists z) (\varphi (z) \land \neg \chi (z)) \Bigg]$ \\
    axiom (FO2)
    \item[65.] $\Bigg[ \Big[ (\exists y) (\varphi (y) \land (x<y)) \land \Big( \recurrent \Big) \Big] \supset$ \\  
    $\spazio$ $\spazio$ $\spazio$ $ \Big( \recurrent \Big) \supset (\exists z) (\varphi (z) \land \neg \chi (z)) \Bigg] \supset$ \\
    $\spazio$ $\spazio$ $\spazio$ $\Bigg[ \Big[ (\exists y) (\varphi (y) \land (x<y)) \land \Big( \recurrent \Big) \Big] \supset (\exists z) (\varphi (z) \land \neg \chi (z)) \Bigg]$ \\
    by (MP) from 63. and 64.
    \item[66.] $\Big[ (\exists y) (\varphi (y) \land (x<y)) \land \Big( \recurrent \Big) \Big] \supset (\exists z) (\varphi (z) \land \neg \chi (z))$ \\
    by (MP) from 62. and 65.
    \item[67.] $\Big( \recurrent \Big) \supset \Big[ (\exists y) (\varphi (y) \land (x<y)) \supset (\exists z) (\varphi (z) \land \neg \chi (z)) \Big]$ \\
    by (R3) from 66.
    \item[68.] $ \Big[ \Big( \recurrent \Big) \supset (\exists y) (\varphi (y) \land (x<y)) \Big] \supset $ \\
    $\spazio$ $\spazio$ $\spazio$ $\Bigg[ \Big( \recurrent \Big) \supset \Big[ (\exists y) (\varphi (y) \land (x<y)) \supset (\exists z) (\varphi (z) \land \neg \chi (z)) \Big] \Bigg] \supset$ \\
    $\spazio$ $\spazio$ $\spazio$ $\spazio$ $\spazio$ $\spazio$ $\Bigg[ \Big[ \Big( \recurrent \Big) \supset (\exists z) (\varphi (z) \land \neg \chi (z)) \Big] \Bigg]$ \\
    axiom (FO2)
    \item[69.] $\Bigg[ \Big( \recurrent \Big) \supset \Big[ (\exists y) (\varphi (y) \land (x<y)) \supset (\exists z) (\varphi (z) \land \neg \chi (z)) \Big] \Bigg] \supset$ \\
    $\spazio$ $\spazio$ $\spazio$ $\Big[ \Big( \recurrent \Big) \supset (\exists z) (\varphi (z) \land \neg \chi (z)) \Big] $ \\
    by (MP) from 9. abd 68.
    \item[70.] $\Big( \recurrent \Big) \supset (\exists z) (\varphi (z) \land \neg \chi (z)) $ \\
    by (MP) from 67. and 69.
    \item[71.] $(\exists x) \Big( \recurrent \Big) \supset (\exists z) (\varphi (z) \land \neg \chi (z)) $ \\ 
    by (R2) from 70.
    \item[72.] $\exists x \Big( \recurrent \Big)$ \\
    axiom (Ax1)
    \item[73.] $(\exists z) (\varphi (z) \land \neg \chi (z))$ \\
    by (MP) from 72. and 71.
\end{itemize}
\end{proof}

\end{document}